\documentclass[aps,prd,showpacs]{revtex4}
\usepackage[dvips]{graphicx}
\newcommand \beq{\begin{eqnarray}}
\newcommand \eeq{\end{eqnarray}}

\def\simge{\mathrel{%
       \rlap{\raise 0.511ex \hbox{$>$}}{\lower 0.511ex \hbox{$\sim$}}}}
\def\simle{\mathrel{
       \rlap{\raise 0.511ex \hbox{$<$}}{\lower 0.511ex \hbox{$\sim$}}}}

\begin{document}

\title{Magnetic vortex in color-flavor locked quark matter}
\author{Kei Iida}
\affiliation{RIKEN BNL Research Center, Brookhaven National Laboratory,
Upton, New York 11973}
\date{\today}

\begin{abstract}
     Within Ginzburg-Landau theory, we study the structure of a magnetic 
vortex in color-flavor locked quark matter.  This vortex is characterized by 
winding of the $SU(3)$ phase in color-flavor space, as well as by the presence
of a color-flavor unlocked condensate in the core.  We estimate the upper 
and lower critical fields and the critical Ginzburg-Landau parameter that
distinguishes between type I and type II superconductors.

\end{abstract}


\maketitle

\section{Introduction}
\label{sec:intro}

     The possibility that dense quark matter occurs in neutron stars has been
considered for the past three decades.  The possible presence of a diquark
condensate in such quark matter \cite{CSC-review} has generated new interest
in such matter.  Among various condensates, a color-flavor locked (CFL) state
\cite{CFL}, in which all three color and flavors are gapped, has been studied
most intensively, along with a two-flavor pairing state \cite{BL}.  In the 
weak coupling (high density) and massless limit the CFL state is most stable 
at zero temperature \cite{HH} and near the transition temperature \cite{I}, a
feature ensured by the quark-quark attraction in the color antitriplet channel
due to one-gluon exchange.  Although the massless, weak coupling results for 
diquark condensation are modified by effects of strong coupling and nonzero 
quark masses in a regime as might be encountered in possible quark matter 
in neutron star interiors \cite{Huang}, the CFL phase has to be reckoned 
with as a reference state in such a regime.  

     Response of the CFL condensate to magnetic fields, as would be
experienced by quark matter if present in neutron stars, was 
considered in Refs.\ \cite{BL,BS,ABR,III,GR}.  Because of nonvanishing color 
and electric charge carried by diquark pairs, electromagnetic superconductivity
and color superconductivity occur in the CFL phase.  These two phenomena are 
not independent in the sense that transverse photon and gluon fields are mixed 
with each other.  One of the resultant mixed fields is freely propagating, 
while the other is Meissner screened and thus massive.  The massive mixed 
field and the winding of the $SU(3)$ phase in color-flavor space are essential
to the structure of a magnetic vortex in the CFL phase.  This point was not 
taken into account in Ref.\ \cite{III}.  The criterion of whether or not the 
CFL state can allow magnetic vortices to form (type II or type I) has been 
described by Giannakis and Ren \cite{GR} in terms of the energy needed to form
a planar surface separating the normal and superconducting material, but the 
detailed structure of the magnetic vortex has yet to be determined.

     In this paper we examine the properties of magnetic vortices in the CFL
phase within the framework of Ginzburg-Landau theory.  These vortices are
characterized by three different fields: the CFL diquark
field, the massive photon-gluon mixed field, and a color-flavor unlocked
diquark field coupled to the mixed field \cite{GR}.  In the London limit where
the wavelengths of distortions of the condensate are long compared with the 
size of the vortex core, the supercurrents surrounding the core can be 
written in terms of the color-flavor $SU(3)$ phase of the CFL condensate and 
the mixed field.  From such supercurrent structures we derive the lower 
critical field, $H_{c1}$.  We next examine the vortex structure near the 
core, in which the color-flavor unlocked condensate plays a role,  
and derive the upper critical field, $H_{c2}$.  Finally we revisit how to 
distinguish between type I and type II superconductors from the viewpoint of
the intervortex interactions.  Throughout this paper, we consider
a system of three-flavor ($uds$) and three-color ($RGB$) massless quarks at 
temperature $T$ and baryon chemical potential $\mu$, and use units 
$\hbar=c=1$.  We assume that the Fermi momentum is common to all colors and 
flavors since in the massless limit of interest here, it is the case with
both the normal and CFL phases of equilibrated neutral quark matter.

\section{Ginzburg-Landau theory}
\label{sec:GL}

     We first review the Ginzburg-Landau theory relevant for the description
of the CFL phase \cite{I,III,GR}.  While this theory is strictly
applicable near the critical temperature $T_c$ and for inhomogeneities 
of wavelengths longer than the zero temperature coherence length $\xi_0
\sim T_c^{-1}$, it is nevertheless useful in describing the structure of a 
possible vortex in the presence of external uniform magnetic fields.  This 
theory focuses only on the pairing channel that first prevails in a normal 
quark-gluon plasma as the temperature goes down.  In the weak coupling 
(high density) regime, this pairing channel is predicted to have even parity, 
zero total angular momentum, and same chirality, and to be in a color and 
flavor antitriplet state.  As we shall see, the CFL phase belongs to this 
pairing channel.  We assume that this channel has the highest transition 
temperature also at lower densities.  This pairing state is characterized by 
a complex $3\times3$ gap matrix, $({\bf d}_a)_i({\bf r})$, in color-flavor 
space \cite{I}, where $a$ ($i$) is the color (flavor) other than two
colors (flavors) involved in Cooper pairing.  This gap is 
defined on the mass shell of the quark quasiparticle of momenta on the Fermi 
surface, but remains dependent on the center-of-mass coordinate ${\bf r}$ of 
the quark Cooper pair.  

     In the presence of an external magnetic field, 
${\bf H}_{\rm ext}=\nabla\times{\bf A}_{\rm ext}$, one can write down the 
thermodynamic potential density difference $\Delta\Omega=\Omega_s-\Omega_n$ 
between the superfluid and normal phases near $T_c$ as \cite{III}
\begin{eqnarray}
   \Delta\Omega= \bar{\alpha} \sum_{a}|{\mathbf{d}}_a|^2
   +\beta_1(\sum_{a}|{\mathbf{d}}_a|^2)^2
   +\beta_2 \sum_{ab}|{\mathbf{d}}_a^{\ast}
    \cdot {\mathbf{d}}_b|^2
   + 2 K_T  \sum_{a}|(D_l {\bf  d})_a|^2
    +\frac{1}{4}G_{lm}^{\alpha}G_{lm}^{\alpha}
    +\frac{1}{4}F_{lm}F_{lm} - \frac{1}{2}|{\bf H}_{\rm ext}|^2.
 \label{GL}
\end{eqnarray}
Here we have neglected the response of the normal component to ${\bf H}_{\rm 
ext}$.  The parameters $\bar\alpha$, $\beta_1$, and $\beta_2$ characterize the
homogeneous part of the free energy, while $K_T$ is the stiffness
parameter, controlling spatial variations of the order parameter.  Since
$({\mathbf{d}}_a)_i$ is antisymmetric in color and flavor space, the covariant
derivative $D_l$ is
\begin{equation}
  (D_l  {\bf d})_a = \partial_l {\bf d}_a
    - \frac{i}{2}g A_l^{\alpha} (\lambda^{\alpha}  {\bf d})_a
    - i e A_l (Q{\bf d})_a,
   \label{cd2}
\end{equation}
where the $\lambda^{\alpha}$ are the Gell-Mann matrices. 
$G_{lm}^{\alpha}$ $(F_{lm})$ is the spatial part of the gluon (photon) 
field-strength tensor, $Q={\rm diag}(2,-1,-1)/3$ is the electric charge 
matrix in flavor space, and $g$ is the QCD coupling constant.
Here we have adopted the convention in which $\lambda^8={\rm diag}(2,-1,-1)
/\sqrt3$ and $({\bf d}_a)_i$ is defined in terms of the adjoint spinors,
rather than the spinors \cite{note1}.

     The parameters $\bar\alpha$, $\beta_1$, $\beta_2$, and $K_T$ 
in Eq.\ (\ref{GL}) can in principle be determined as functions of $T$ 
and $\mu$.  In a general Ginzburg-Landau framework, however, these
parameters are left unknown.  The parameters have been determined only in the 
weak coupling limit in which the pairing interaction is controlled by
Landau-damped magnetic gluons, and are \cite{I,GR1}
\begin{eqnarray}
{\bar\alpha}=4N(\mu/3)\ln\left(\frac{T}{T_{c}}\right),  \\
\beta_{1}=\beta_{2}= 3K_T=\frac{7\zeta(3)}{8(\pi T_c)^{2}}
   N(\mu/3),\\
 N(\mu/3) =\frac{1}{2\pi^{2}}\left(\frac{\mu}{3}\right)^{2},
\end{eqnarray}
with the zeta function $\zeta(3)=1.202\cdots$.

     In homogeneous quark matter in weak coupling, the energy difference 
(\ref{GL}) has a minimal value \cite{I} just below $T_c$ for a CFL order 
parameter, i.e.,
\begin{equation}
   ({\bf d}_a)_i=U_{ai}\kappa_A, ~~\kappa_A=e^{i\varphi_0}|\kappa_A|,
  \label{cfl}
\end{equation}
where $U\equiv \exp(i\lambda^\alpha \varphi_\alpha /2)$ 
represents $SU(3)$ rotations in color-flavor space.  Generally,
in a homogeneous situation, the CFL state is the most stable as long as 
$3\beta_1+\beta_2>0$ and $\beta_2>0$ \cite{I}.  In terms of $\bar\alpha$,
$\beta_1$, and $\beta_2$, the minimizing value of $|\kappa_A|$ is
\begin{equation}
   |\kappa_A|=\left(\frac{-\bar\alpha}{6{\bar\beta}_{\rm CFL}}\right)^{1/2},
\end{equation}
where ${\bar\beta}_{\rm CFL}=\beta_1+\beta_2/3$.

     We can now derive the field equations characterizing the CFL
state under constant external magnetic field, ${\bf H}_{\rm ext}$.  In 
this situation, as clarified in Ref.\ \cite{GR}, the electromagnetic field,
which is mixed with the gluon field of $\alpha=8$, is coupled with the
color-flavor rotation $\exp(i\lambda^8 \varphi_8 /2)$ of the gap matrix
$({\bf d}_a)_i$.  Interestingly, the solution to the field equations requires
a nonvanishing color-flavor unlocked diquark component proportional to 
$[\lambda^8 \exp(i\lambda^8 \varphi_8 /2)]_{ai}$, in addition to the CFL 
component proportional to $[\exp(i\lambda^8 \varphi_8 /2)]_{ai}$ \cite{GR};  
addition of any other components would simply increase the total free energy
of the system.  It is thus convenient to introduce the ansatz for the pairing 
gap \cite{GR}: 
\begin{eqnarray}
   ({\bf d}_a)_i({\bf r})&=&\delta_{aR}\delta_{iu}\phi({\bf r})
                    +(\delta_{aG}\delta_{id}+\delta_{aB}\delta_{is})
                     \chi({\bf r})/\sqrt2
  \nonumber \\
    &=& [e^{i\lambda^8 \varphi_8 ({\bf r}) /2}]_{ai} \frac{|\phi({\bf r})|
        +\sqrt2|\chi({\bf r})|}{3}
       +\left[\frac{\lambda_8}{2} 
               e^{i\lambda^8 \varphi_8 ({\bf r}) /2}\right]_{ai} 
        \frac{2|\phi({\bf r})|-\sqrt2|\chi({\bf r})|}{\sqrt3},
  \label{cfl2}
\end{eqnarray}
with $\phi({\bf r})=\exp[i\varphi_8({\bf r)}/\sqrt3]|\phi({\bf r})|$ and 
$\chi({\bf r})=\exp[-i\varphi_8({\bf r)}/2\sqrt3]|\chi({\bf r})|$,
and the photon-gluon mixed fields \cite{gorbar}:
\begin{equation}
   \mbox{\boldmath ${\cal A}$}
   \equiv \frac{\sqrt3 g {\bf A} - 2e {\bf A}^8}{6g_{\rm CFL}}, ~~
   \mbox{\boldmath ${\cal A}$}^8
   \equiv \frac{\sqrt3 g {\bf A}^8 + 2e {\bf A}}{6g_{\rm CFL}},
  \label{cflmix}
\end{equation}
where $g_{\rm CFL}=\sqrt{3g^2+4e^2}/6$.  Then, we can express the 
thermodynamic potential density difference as
\begin{eqnarray}
    \Delta\Omega&=&{\bar\alpha}(|\phi|^2+|\chi|^2)
    +\beta_1(|\phi|^2+|\chi|^2)^2+\beta_2(|\phi|^4+|\chi|^4/2)
   \nonumber \\  & &
    +2K_T\left|\left(\nabla- 2 i g_{\rm CFL} 
         \mbox{\boldmath ${\cal A}$}^8\right)\phi\right|^2
    +2K_T\left|\left(\nabla+i g_{\rm CFL} 
         \mbox{\boldmath ${\cal A}$}^8\right)\chi\right|^2
    +\frac{1}{2}(|\mbox{\boldmath ${\cal B}$}|^2
                 +|\mbox{\boldmath ${\cal B}$}^8|^2
                 -|{\bf H}_{\rm ext}|^2),
   \label{GL2}
\end{eqnarray}
where $\mbox{\boldmath ${\cal B}$}=\nabla\times\mbox{\boldmath ${\cal A}$}$
and $\mbox{\boldmath ${\cal B}$}^8=\nabla\times\mbox{\boldmath ${\cal A}$}^8$.
By extremizing this free energy difference with respect to the pairing gaps
and gauge fields, we obtain the gap equations,
\begin{equation}
  -2K_T(\nabla-2ig_{\rm CFL}\mbox{\boldmath ${\cal A}$}^8)^2\phi
   +{\bar\alpha}\phi+2\beta_1(|\phi|^2+|\chi|^2)\phi
   +2\beta_2 |\phi|^2 \phi=0,
   \label{gephi}
\end{equation}
\begin{equation}
  -2K_T\left(\nabla+ig_{\rm CFL}\mbox{\boldmath ${\cal A}$}^8\right)^2\chi
   +{\bar\alpha}\chi+2\beta_1(|\phi|^2+|\chi|^2)\chi
   +\beta_2 |\chi|^2 \chi=0,
   \label{gechi}
\end{equation}
and the classical Maxwell equations for the photon-gluon mixed fields,
\begin{eqnarray}
  \nabla\times(\nabla\times\mbox{\boldmath ${\cal A}$}^8)
  &=&{\rm Re}\{4i K_T g_{\rm CFL} 
 [-2\phi^*(\nabla-2ig_{\rm CFL} \mbox{\boldmath ${\cal A}$}^8)\phi
  +\chi^*(\nabla+ig_{\rm CFL} \mbox{\boldmath ${\cal A}$}^8)\chi]\}
  \nonumber \\ 
 &\equiv&\mbox{\boldmath ${\cal J}$}^8,
   \label{max8}
\end{eqnarray}
\begin{equation}
  \nabla\times(\nabla\times\mbox{\boldmath ${\cal A}$})=0.
   \label{max}
\end{equation}
From Eqs.\ (\ref{max8}) and (\ref{max}) we find that in the CFL state
($|\phi|=|\chi|/\sqrt2=|\kappa_A|$), the mixed field
$\mbox{\boldmath ${\cal A}$}^8$ is Meissner screened within a London
penetration depth,
\begin{equation}
   \lambda_{\rm CFL}=\frac{1}{2\sqrt{6 K_T} g_{\rm CFL} |\kappa_A|},
   \label{london}
\end{equation}
whereas the mixed field $\mbox{\boldmath ${\cal A}$}$ is freely propagating.
     
     The thermodynamic critical field, $H_c$, associated with ${\bf B}=
\nabla\times{\bf A}$ is the field at which the Gibbs free energy density of 
the normal state, $G_n=\Omega_n-{\bf H}_{\rm ext}\cdot{\bf B}$, drops to that 
of the superconducting state, $G_s=\Omega_s-{\bf H}_{\rm ext}\cdot{\bf B}$.  
Since $\mbox{\boldmath ${\cal B}$}^8$ is screened out in the CFL phase in 
bulk, one can obtain the Gibbs free energy difference as \cite{III}
\begin{equation}
  \Delta G \equiv G_s-G_n =-\frac{{\bar\alpha}^2}{4{\bar\beta}_{\rm CFL}}
                 +\frac{e^2}{18g_{\rm CFL}^2}|{\bf H}_{\rm ext}|^2.
\end{equation}
Then, 
\begin{equation}
   H_c=\frac{3}{2e\xi_{\rm CFL}\lambda_{\rm CFL}},
    \label{hc}
\end{equation}
where 
\begin{equation}
 \xi_{\rm CFL}=\left(\frac{2K_T}{|{\bar\alpha}|}\right)^{1/2}
    \label{xi}
\end{equation}
is the Ginzburg-Landau coherence length.  As discussed in Ref.\ \cite{III},
extrapolation of the weak coupling expression for $H_c$ to low densities and 
temperatures indicates that the critical field $H_c$ is typically 
$\sim10^{19}$ G, several orders of magnitude larger than canonical neutron 
star surface fields $\sim10^{12}$ G.

     We conclude this section by asking whether or not the CFL condensate 
allows magnetic vortices associated with the field 
$\mbox{\boldmath ${\cal A}$}^8$ to form, i.e., whether it is type II or
type I.  In a standard method for distinguishing between type I and type II 
superconductors, one calculates the energy per unit area, $\sigma_s$, needed
to form a planar surface separating the normal and superconducting material.
At the thermodynamic critical field, (\ref{hc}), applied parallel to 
the surface, the surface is in mechanical equilibrium.  For a surface 
perpendicular to the $z$ axis, the surface energy can be written as the
integral over $z$ of the difference between the total Gibbs free energy 
density and the value of $|z|\to\infty$ \cite{III}:
\begin{equation}
     \sigma_s=\int_{-\infty}^{\infty}dz[\Delta\Omega(z)|_{H_{\rm ext}\to
     H_c}-H_c(|{\bf B}(z)|-H_c)].
\end{equation}
Because of the presence of the color-flavor unlocked component near the
surface, the system remains type II, i.e., $\sigma_s <0$,
even when the Ginzburg-Landau parameter, 
$\kappa_{\rm CFL}\equiv\lambda_{\rm CFL}/\xi_{\rm CFL}$, becomes less than 
$1/\sqrt2$ by a small amount \cite{GR}.  This is a contrast to the case of 
ordinary superconductors in which the Ginzburg-Landau parameter is $1/\sqrt2$
when the interfacial energy vanishes \cite{dG}.  Detailed calculations 
\cite{GR} show that for $\beta_1=\beta_2$ the system remains type II for 
values of $\kappa_{\rm CFL}$ down to $\simeq0.589$.  As we shall see in Sec.\ 
\ref{sec:struc}, however, the properties of intervortex interactions imply a 
significantly narrower type II region.

\section{Magnetic response in the London limit}
\label{sec:london}

     We now examine the response of the CFL condensate to external magnetic 
fields by deriving the supercurrent $\mbox{\boldmath ${\cal J}$}^8$ in the 
London limit in which a length scale of the spatial variation of the 
condensate is large compared with the coherence length $\xi_{\rm GL}$, 
corresponding to the core size of a magnetic vortex.  In this limit, one can 
ignore the variation of the magnitude of the pairing gaps, $\phi$ and $\chi$, 
as compared with the variation of their phases \cite{III}.  We may thus derive
$\mbox{\boldmath ${\cal J}$}^8$ by substituting the ansatz (\ref{cfl2}) with 
$|\phi({\bf r})|=|\chi({\bf r})|/\sqrt2=|\kappa_A|$ into Eq.\ (\ref{max8}).
The result reads 
\begin{equation}
 \mbox{\boldmath ${\cal J}$}^8=4K_T g_{\rm CFL}|\kappa_A|^2
      (\sqrt3\nabla\varphi_8-6g_{\rm CFL}\mbox{\boldmath ${\cal A}$}^8).
   \label{j8}
\end{equation}
We thus find that this supercurrent can be induced by the gradient of the 
color-flavor $SU(3)$ phase, $\varphi_8$.  This feature is essential to 
magnetic vortices, but was ignored in Ref.\ \cite{III}.  From expression 
(\ref{j8}) we can examine the response properties of the CFL condensate.

     We first consider the response to weak uniform magnetic fields.  As
discussed in Ref.\ \cite{III}, the response is characterized by imperfect 
diamagnetism, i.e., partial Meissner screening.  As long as $g\gg e$, most
of the field freely propagates in the form of 
$\mbox{\boldmath ${\cal B}$}$, whereas the rest is included in 
$\mbox{\boldmath ${\cal B}$}^8$ and hence screened on a length scale of 
$\lambda_{\rm CFL}$.  

     For high fields, it is interesting to consider the possible presence of
magnetic vortices.  We find from Eq.\ (\ref{j8}) that vortices can appear in 
such a way as to satisfy the flux quantization condition,
\begin{equation} 
 \oint d{\bf \ell}\cdot (\mbox{\boldmath ${\cal A}$}^8
                 +\lambda_{\rm CFL}^2\mbox{\boldmath ${\cal J}$}^8)
 =\frac{2\pi n}{g_{\rm CFL}}\equiv\phi_8 n,
  \label{quan}
\end{equation} 
where the integration is performed around a closed loop surrounding the vortex
line, $\phi_8$ is the flux quantum, and $n$ is the winding number of the
vortex.  Whether vortices actually occur in equilibrium depends on the value 
of $\kappa_{\rm CFL}$, as will be discussed in Sec.\ \ref{sec:struc}.

     In the London limit we can analyze the structure, far away from 
the core, of a singly quantized ($n=1$) vortex and then estimate the lower 
critical field, $H_{c1}$, above which a single vortex starts to appear.  Let 
us take the vortex to be aligned along the $z$ axis and set 
$\varphi_8=2\sqrt3\varphi$, where $\varphi$ is the azimuthal angle around the 
line.  Then, the field equation (\ref{max8}) reduces to the London equation,
\begin{equation}
   \mbox{\boldmath ${\cal B}$}^8 + \lambda_{\rm CFL}^2 \nabla\times
    (\nabla\times \mbox{\boldmath ${\cal B}$}^8) =
     \frac{1}{2\sqrt3 g_{\rm CFL}}\nabla\times\nabla\varphi_8
    = \frac{2\pi}{g_{\rm CFL}}\delta(x)\delta(y) {\hat z}.
  \label{max8lon}
\end{equation}
Note that this equation leads to Eq.\ (\ref{quan}) with $n=1$.

      As in ordinary type II superconductors \cite{dG}, one can solve the
London equation (\ref{max8lon}) and obtain the supercurrent density flowing 
in the azimuthal direction around the line as
\begin{equation}
  \mbox{\boldmath ${\cal J}$}^8 =
 \left\{
 \begin{array}{ll}
  \displaystyle{\frac{\phi_8}{2\pi\lambda_{\rm CFL}^2 r}}{\hat \varphi},
   & \quad \mbox{for $\xi_{\rm CFL} \ll r \ll \lambda_{\rm CFL}$}, \\
  \displaystyle{\frac{\phi_8}{2\pi\lambda_{\rm CFL}^2}
                \left(\frac{\pi\lambda_{\rm CFL}}{2r}\right)^{1/2}
      {\hat \varphi}
          \left[\frac{1}{\lambda_{\rm CFL}}
             +\frac{1}{2r}
             +{\cal O}\left(\frac{\lambda_{\rm CFL}}{r^2}\right)\right]}
             e^{-r/\lambda_{\rm CFL}},
   & \quad \mbox{for $r \gg \lambda_{\rm CFL}$}. \\
 \end{array}
 \right.
     \label{curlon}
\end{equation}
The decrease of this supercurrent at a scale of $\lambda_{\rm CFL}$
suggests that the field $\mbox{\boldmath ${\cal B}$}^8$ associated with the
vortex can penetrate only up to such a scale, beyond which 
$\mbox{\boldmath ${\cal A}$}^8 \simeq (1/g_{\rm CFL}r){\hat \varphi}$ 
and thus $\mbox{\boldmath ${\cal B}$}^8 \simeq0$.  Combining 
Eqs.\ (\ref{quan}) and (\ref{curlon}), one can 
calculate the vortex line energy per unit length, which is composed of the 
sum of the magnetic and flow energies, as 
\begin{equation}
 T_L = \frac{\phi_8^2}{4\pi\lambda_{\rm CFL}^2}
     \ln\kappa_{\rm CFL}.
  \label{tlcfl}
\end{equation}

     At the lower critical field $H_{c1}$, the line energy is balanced by 
the energy gain due to the magnetic induction 
$\mbox{\boldmath ${\cal B}$}^8 = {\cal B}^8_z {\hat z}$.
The energy gain per unit volume is $E_{\rm mag}=-{\cal B}^8_z {\cal H}^8$, 
where ${\cal H}^8=(e/3g_{\rm CFL})H_{c1}$ is the critical field associated with
the massive photon-gluon mixed field [see Eq.\ (\ref{cflmix})].  Integrating
over the volume including the vortex, one can estimate a total energy gain
per unit length as $-\phi_8(e/3g_{\rm CFL})H_{c1}$.  Then,
\begin{equation}
 H_{c1} = \frac{3}{2e\lambda_{\rm CFL}^2}
          \ln\kappa_{\rm CFL}.
   \label{hc1}
\end{equation}
Note that the logarithmic factor in this expression contains uncertainties 
of order unity.

\section{Structure of magnetic vortices}
\label{sec:struc}

     As shown in the previous section, the analyses in the London limit are 
useful in clarifying the global properties of the magnetic response in the CFL
phase.  However, it is important to examine the vortex structure near the 
core, e.g., in evaluating the upper critical field $H_{c2}$ where the
vortices essentially fuse and the system becomes normal.  In this section we 
thus go back to the Ginzburg-Landau equations and examine their solutions.
Using these results, we revisit the problem of how to distinguish between 
type I and type II superconductors.

     It is useful to start with the linearized gap equations 
with respect to $\phi$ and $\chi$:
\begin{equation}
  -2K_T(\nabla-2ig_{\rm CFL}\mbox{\boldmath ${\cal A}$}^8)^2\phi
   +{\bar\alpha}\phi=0,
   \label{gephilin}
\end{equation}
\begin{equation}
  -2K_T\left(\nabla+ig_{\rm CFL}\mbox{\boldmath ${\cal A}$}^8\right)^2\chi
   +{\bar\alpha}\chi=0.
   \label{gechilin}
\end{equation}
This is because equations (\ref{gephilin}) and (\ref{gechilin}) can describe 
the situation near $H_{c2}$ in which the pairing gaps are suppressed in
magnitude.  The critical condition for the existence of nontrivial 
solutions to at least one of these equations is satisfied when 
$g_{\rm CFL}|\mbox{\boldmath ${\cal B}$}^8|=|{\bar\alpha}|/2K_T$.  In 
this situation, a nontrivial $\chi$ solution to Eq.\ (\ref{gechilin}) occurs, 
while $\phi=0$.  Using Eq.\ (\ref{cflmix}) we thus obtain
\begin{equation}
    H_{c2}=\frac{3}{e\xi_{\rm CFL}^2}.     
   \label{hc2}
\end{equation}

     Combining Eqs.\ (\ref{hc}) and (\ref{hc2}), we find that
$H_{c2}/H_c=2\kappa_{\rm CFL}$.  When $H_{c2}>H_c$, i.e., $\kappa_{\rm CFL}>
1/2$, magnetic vortices are expected to occur.  This condition in fact 
disagrees with the criterion for type II superconductors as derived in Ref.\ 
\cite{GR} from the sign of the normal-super interfacial energy $\sigma_s$ (see
Sec.\ \ref{sec:GL}).  Note that in ordinary superconductors \cite{dG}, the 
condition $H_{c2}=H_c$ agrees with the condition $\sigma_s=0$.  The difference
between these two cases arises from the fact that in the CFL case, magnetic 
vortices are characterized by the color-flavor locked and unlocked 
condensates, while in the ordinary case, they are characterized by only a 
single complex scalar condensate. 

     This two-field nature of the CFL vortices can be seen by substituting
$\phi(r)=|\kappa_A|\exp(2in\varphi)u(r)$, $\chi(r)=\sqrt2|\kappa_A|
\exp(-in\varphi)v(r)$, and $\mbox{\boldmath ${\cal A}$}^8(r)=
{\cal A}_\varphi^8(r) {\hat \varphi}$ into the field
equations (\ref{gephi}), (\ref{gechi}), and (\ref{max8}).  Then we obtain 
\begin{equation}
  -2K_T\left(\frac{d^2}{dr^2}+\frac{1}{r}\frac{d}{dr}\right)u
  +8K_T\left(\frac{n}{r}-g_{\rm CFL}{\cal A}_\varphi^8\right)^2 u
  +{\bar\alpha}u+2\beta_1 |\kappa_A|^2 (u^2+2v^2)u
   +2\beta_2 |\kappa_A|^2 u^3=0,
   \label{geu}
\end{equation}
\begin{equation}
  -2K_T\left(\frac{d^2}{dr^2}+\frac{1}{r}\frac{d}{dr}\right)v
  +2K_T\left(\frac{n}{r}-g_{\rm CFL}{\cal A}_\varphi^8\right)^2 v
  +{\bar\alpha}v+2\beta_1 |\kappa_A|^2 (u^2+2v^2)v
   +2\beta_2 |\kappa_A|^2 v^3=0,
   \label{gev}
\end{equation}
\begin{eqnarray}
  -\frac{d}{dr}\frac{1}{r}\frac{d}{dr}r{\cal A}_\varphi^8
  = \frac{1}{3g_{\rm CFL}\lambda_{\rm CFL}^2} 
         (2u^2+v^2)
    \left(\frac{n}{r}-g_{\rm CFL}{\cal A}_\varphi^8\right).
   \label{maxphi8}
\end{eqnarray}

     In the limit $r\to0$, the normalized gaps $u$ and $v$ and the gauge 
field ${\cal A}_\varphi^8$ are suppressed in magnitude.  Consequently, the 
gap equations reduce to
\begin{equation}
  -\left(\frac{d^2}{dr^2}+\frac{1}{r}\frac{d}{dr}\right)u
  +4\left(\frac{n}{r}\right)^2 u \to 0,
   \label{geu0}
\end{equation}
\begin{equation}
  -\left(\frac{d^2}{dr^2}+\frac{1}{r}\frac{d}{dr}\right)v
  +\left(\frac{n}{r}\right)^2 v \to 0.
   \label{gev0}
\end{equation}
From Eqs.\ (\ref{geu0}) and (\ref{gev0}), we find that 
\begin{equation}
    u\to C_1 r^{2n},~~~ v\to C_2 r^n, 
\end{equation}
where $C_1$ and $C_2$ are real constants that can be
determined by connecting the solution near the center ($r=0$) with that 
outside the core.  We can then calculate the asymptotic behavior of 
${\cal A}_\varphi^8$ from
\begin{eqnarray}
  -\frac{d}{dr}\frac{1}{r}\frac{d}{dr}r{\cal A}_\varphi^8
  \to \frac{1}{3g_{\rm CFL}\lambda_{\rm CFL}^2} 
      (2u^2+v^2)\frac{n}{r};
   \label{maxphi80}
\end{eqnarray}
the result reads 
\begin{equation}
{\cal A}_\varphi^8 \to \frac12 {\cal B}^8_0 r
   -\frac{C_2^2}{12g_{\rm CFL}\lambda_{\rm CFL}^2 (n+1)} 
     r^{2n+1},
\end{equation}
where ${\cal B}^8_0={\cal B}^8_z (r=0)$.  Consequently,
\begin{equation}
{\cal B}_z^8 \to {\cal B}^8_0
   -\frac{C_2^2}{6g_{\rm CFL}\lambda_{\rm CFL}^2} r^{2n}.
\end{equation}
We remark that as $r\to0$, $v$ dominates over $u$.  From Eq.\ (\ref{cfl2}) we 
thus find that the color-flavor locked and unlocked condensates coexist
in the vortex core and behave as $r^n$ near the center.  This is
similar to the case near $H_{c2}$ in which $\phi$ essentially vanishes.

     We proceed to examine the vortex structure outside the core 
($r \gg \xi_{\rm CFL}$) by taking into account small deviations from the 
asymptotic behaviors discussed in Sec.\ \ref{sec:london}: 
$\delta u(r)=u(r)-1$, $\delta v(r)=v(r)-1$, and 
$\delta{\cal A}_\varphi^8(r)={\cal A}_\varphi^8(r)-n/g_{\rm CFL}r$.
The field equations of linear order in these deviations read
\begin{equation}
  -2K_T\left(\frac{d^2}{dr^2}+\frac{1}{r}\frac{d}{dr}\right)\delta u
  +[{\bar\alpha}+(10\beta_1+6\beta_2)|\kappa_A|^2]\delta u 
  +8\beta_1|\kappa_A|^2 \delta v =0,
   \label{deltau}
\end{equation}
\begin{equation}
  -2K_T\left(\frac{d^2}{dr^2}+\frac{1}{r}\frac{d}{dr}\right)\delta v
  +[{\bar\alpha}+(14\beta_1+6\beta_2)|\kappa_A|^2]\delta v
  +4\beta_1|\kappa_A|^2 \delta u =0,
   \label{deltav}
\end{equation}
\begin{eqnarray}
  \frac{d}{dr}\frac{1}{r}\frac{d}{dr}r \delta{\cal A}_\varphi^8
  = \frac{1}{\lambda_{\rm CFL}^2}\delta{\cal A}_\varphi^8.
   \label{deltaa8}
\end{eqnarray}
It is convenient to diagonalize Eqs.\ (\ref{deltau}) and (\ref{deltav}) as
\begin{equation}
  \left(\frac{d^2}{dr^2}+\frac{1}{r}\frac{d}{dr}\right)
   \frac{\delta u +2\delta v}{3}
  =\frac{2}{\xi_{\rm CFL}^2}
   \frac{\delta u +2\delta v}{3},
   \label{deltacfl}
\end{equation}
\begin{equation}
  \left(\frac{d^2}{dr^2}+\frac{1}{r}\frac{d}{dr}\right)
   \frac{2(\delta u -\delta v)}{\sqrt3}
  =\frac{2}{\epsilon\xi_{\rm CFL}^2}
   \frac{2(\delta u - \delta v)}{\sqrt3},
   \label{deltacful}
\end{equation}
where $\epsilon=3{\bar\beta}_{\rm CFL}/\beta_2$.  Note that
$(\delta u+2\delta v)/3$ represents a change in magnitude of 
the CFL condensate from the homogeneous one, whereas 
$2(\delta u- \delta v)/\sqrt3$ 
represents the magnitude of the color-flavor unlocked condensate
[see Eq.\ (\ref{cfl2})].  We thus find from Eqs.\ (\ref{deltacfl}) and 
(\ref{deltacful}) as well as the $r\to 0$ behaviors discussed above that
the magnitude of the CFL condensate increases with $r$ in proportion to
$r^n$ near the vortex center and, above a scale of $\xi_{\rm CFL}/\sqrt2$, 
approaches exponentially the homogeneous solution ($u=v=1$); the 
magnitude of the color-flavor unlocked condensate increases with $r$
in proportion to $r^n$ near the vortex center and, above a scale of 
$\sqrt{\epsilon/2}\xi_{\rm CFL}$, decreases exponentially to zero.  We remark 
that for $r\gg\lambda_{\rm CFL}$, $\delta{\cal A}_\varphi^8$ follows the 
London limit behavior of the supercurrent (\ref{curlon}) via 
$\delta{\cal A}_\varphi^8 = -n\lambda_{\rm CFL}^2
|\mbox{\boldmath ${\cal J}$}^8|$, leading to 
${\cal B}_z^8 \simeq (n\phi_8 / 2\pi\lambda_{\rm CFL}^2)
(\pi\lambda_{\rm CFL}/2r)^{1/2}\exp(-r/\lambda_{\rm CFL})$.
The vortex structure thus clarified is schematically illustrated in Fig.\ 1.

\begin{figure}[t]
\begin{center}
\includegraphics[width=8.5cm]{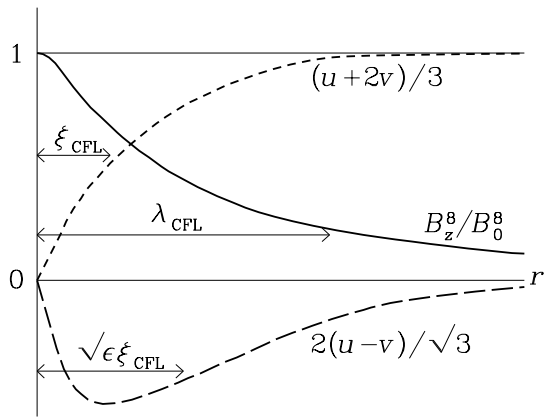}
\end{center}
\vspace{-0.5cm}
\caption{
Schematic variation of the mixed field ${\cal B}_z^8$ (solid line),
the CFL condensate $(u+2v)/3$ (dotted line), and the 
color-flavor unlocked condensate $2(u-v)/\sqrt3$ (dashed line)
for a singly quantized vortex 
($\lambda_{\rm CFL} > \sqrt\epsilon \xi_{\rm CFL} > \xi_{\rm CFL}$).
}
\end{figure}

     The parameter $\epsilon=3{\bar\beta}_{\rm CFL}/\beta_2$ characterizing 
the scale of the spatial variation of the color-flavor unlocked condensate 
plays a important role in determining the nature of the interaction between 
widely separated vortices.  For the purpose of estimating the intervortex 
interaction, it is instructive to write down the solutions to Eqs.\ 
(\ref{deltaa8}), (\ref{deltacfl}), and (\ref{deltacful}) as
\begin{equation}
 \ln(g_{\rm CFL}\lambda_{\rm CFL}\delta{\cal A}_\varphi^8) \simeq 
 \ln\left[K_1\left(\frac{r}{\lambda_{\rm CFL}}\right)
   \right] + {\cal O}(1),
\end{equation}
\begin{equation}
 \ln \frac{\delta u+ 2\delta v}{3} \simeq \ln\left[K_0
  \left(\frac{\sqrt2 r}{\xi_{\rm CFL}}\right)\right] + {\cal O}(1),
\end{equation}
\begin{equation}
 \ln \frac{2(\delta u-\delta v)}{\sqrt3} \simeq \ln\left[K_0\left(
   \frac{\sqrt2 r}{\sqrt\epsilon \xi_{\rm CFL}}\right)\right] + {\cal O}(1),
\end{equation}
where $K_m$ are the $m$th modified Bessel functions.  Then, following
Ref.\ \cite{mackenzie}, one can estimate the interaction energy per unit 
length of two vortices at large separation $a_v$ as
\begin{equation}
  U(a_v) \simeq K_T|\kappa_A|^2 \left[
        C_3 K_0\left(\frac{a_v}{\lambda_{\rm CFL}}\right)
       -C_4 K_0\left(\frac{\sqrt2 a_v}{\xi_{\rm CFL}}\right)
       -C_5 K_0\left(\frac{\sqrt2 a_v}{\sqrt{\epsilon}\xi_{\rm CFL}}\right)
                  \right],
   \label{int}
\end{equation}
where $C_3$, $C_4$, and $C_5$ are positive dimensionless constants.  Since 
$K_{0}(x)$ is positive definite and decreases with $x$, the term associated 
with $\lambda_{\rm CFL}$ in Eq.\ (\ref{int}) induces repulsion between the
vortices, while the terms associated with $\xi_{\rm CFL}$ induce attraction.
This reflects the fact that the total magnetic energy of the two vortices 
becomes larger for smaller intervortex separation, while the total loss
of the condensation energy arising from the two vortices becomes smaller.

     One of the conditions for type II (I) superconductors is that the
intervortex interaction is repulsive (attractive) at large separation.  
In the case of repulsion, a vortex lattice is stable, while in the case of 
attraction, widely separated vortices tend to approach each other more and 
more closely.  From Eq.\ (\ref{int}), whose sign is dominated by the 
exponential behavior of the $K_0$ functions [$\ln K_0(x)= -x+{\cal O}(\ln x)$ 
for large $x$], we find that the intervortex interaction is repulsive when 
$\kappa_{\rm CFL}>{\rm max}(1,\sqrt{\epsilon})/\sqrt2$.  We thus see the 
role played by $\epsilon$ in determining the nature of the intervortex 
interaction.  Note that $\epsilon=4$ in weak coupling.

     The type II region is most strongly restricted by the condition that
the intervortex interaction is repulsive.  In this case, the critical 
value of $\kappa_{\rm CFL}$ is equal to or even larger than $1/\sqrt2$, while 
the value of $\kappa_{\rm CFL}$ at $\sigma_s=0$ and that at $H_c=H_{c2}$ are 
smaller than $1/\sqrt2$.  This is a contrast to the case of ordinary 
superconductors in which, owing to the simple order-parameter structure, all 
three critical values of the Ginzburg-Landau parameter are degenerate at 
$1/\sqrt2$.  We remark that in weak coupling, where $\kappa_{\rm CFL}=
\sqrt{6/7\zeta(3)}(6\pi^2 T_c/g_{\rm CFL}\mu) \ll \sqrt2$ \cite{GR}, 
the CFL condensate near $T_c$ is deep within the type I region.  However, it 
is still uncertain whether it is type I or type II at low densities where 
$T_c$ can be $\sim0.1 \mu$ and $g_{\rm CFL}\sim1$ \cite{III}.  This 
uncertainty affects the estimates of $H_{c1}$ and $H_{c2}$ since 
$H_{c2}=2\kappa_{\rm CFL}H_c$ and $H_{c1}=(\kappa_{\rm CFL}^{-1}
\ln\kappa_{\rm CFL}) H_c$.

\section{Conclusion}
\label{sec:conc}

    Within Ginzburg-Landau theory, we have derived the lower and upper 
critical fields, $H_{c1}$ and $H_{c2}$, for the CFL state, as well as the 
critical Ginzburg-Landau parameter that distinguishes type I and 
type II superconductors.  In doing so, we have examined the structure of a 
magnetic vortex near the center and outside the core.  This vortex is 
associated with the winding of the $SU(3)$ phase in color-flavor 
space, rather than the $U(1)$ electromagnetic phase, and hence allows the 
presence of a color-flavor unlocked condensate in the core.  We have  
found that the nature of the intervortex interaction, controlled by the 
color-flavor unlocked condensate, provides a stringent constraint on the 
type II regime.

     Application of the present analyses to the magnetic structure and 
evolution of neutron stars is not straightforward.  Not only does it require
the extrapolation of the weak coupling results near $T_c$ to low densities and
temperatures, but also the clarification of a role of nonzero quark masses and 
neutrality of electric and color charge in modifying the CFL condensate at
finite temperatures \cite{dSC,RSR,FKR}.  Due to uncertainties associated with 
the extrapolation, it is still open whether the CFL condensate, if occurring 
in stars, would be type I or type II.  How nonzero quark masses and charge 
neutrality affect the response to magnetic fields is another open problem.
Moreover, as discussed in Ref.\ \cite{III}, to assess the actual situations 
possible in neutron stars one must take into account the history of the 
expulsion of the magnetic field and the possibility of freezing in of the 
magnetic field.

\section*{Acknowledgments}

     We thank Gordon Baym for critical reading of the original manuscript
and Ioannis Giannakis and Hai-cang Ren for helpful comments.  We are grateful 
to the Institute for Nuclear Theory at the
University of Washington for its hospitality and to the
Department of Energy for partial support during the completion of this work.


\begin{thebibliography}{99}
    \bibitem{CSC-review} See, e.g., K. Rajagopal and F. Wilczek,
in {\it At the Frontier of Particle Physics, Handbook of QCD, Boris Ioffe
Festschrift, V. 3,} edited by M. Shifman (World Scientific, Singapore, 2001),
p.\ 2061; M.G. Alford, Annu.\ Rev.\ Nucl.\ Part.\ Sci.\ {\bf 51}, 131 (2001).

    \bibitem{CFL} M. Alford, K. Rajagopal, and F. Wilczek, Nucl.\ Phys.\ {\bf
B537}, 443 (1999).

    \bibitem{BL} D. Bailin and A. Love, Phys.\ Rep.\ {\bf 107}, 325 (1984).


    \bibitem{HH} D.K. Hong and S.D.H. Hsu, Phys.\ Rev.\ D {\bf 68}, 034011 
(2003).

    \bibitem{I} K. Iida and G. Baym, Phys.\ Rev.\ D {\bf 63}, 074018 (2001);
Phys.\ Rev.\ D {\bf 66}, 059903(E) (2002).

     \bibitem{Huang} See, e.g., M. Huang, hep-ph/0409167.

    \bibitem{BS} D. Blaschke and D. Sedrakian, nucl-th/0006038.

     \bibitem{ABR} M. Alford, J. Berges, and K. Rajagopal, Nucl.\ Phys.\
{\bf B571}, 269 (2000).

    \bibitem{III} K. Iida and G. Baym, Phys.\  Rev.\  D {\bf 66}, 014015
(2002).

     \bibitem{GR} I. Giannakis and H.-C. Ren, Nucl.\ Phys.\ {\bf B669}, 462
(2003).

\bibitem{note1}
Note the difference from the convention in Ref.\ \cite{III} in which 
$\lambda^8={\rm diag}(1,1,-2)/\sqrt3$.  The present convention is
useful since the corresponding color part in the covariant derivative
is proportional to the electromagnetic part.
    
     \bibitem{GR1} I. Giannakis and H.-C. Ren, Phys.\ Rev.\ D {\bf 65}, 
054017 (2002).

     \bibitem{gorbar} E.V. Gorbar, Phys.\ Rev.\ D {\bf 62}, 014007 (2000).

     \bibitem{dG} P.G. de Gennes, {\it Superconductivity of Metals and 
Alloys} (Benjamin, New York, 1966).

     \bibitem{mackenzie} R. MacKenzie, M.-A. Vachon, and U.F. Wichoski,
Phys.\ Rev.\ D {\bf 67}, 105024 (2003).

     \bibitem{dSC} K. Iida, T. Matsuura, M. Tachibana, and T. Hatsuda,
Phys.\ Rev.\ Lett.\ {\bf 93}, 132001 (2004); hep-ph/0411356.

     \bibitem{RSR} S.B. R{\" u}ster, I.A. Shovkovy, and D.H. Rischke,
Nucl.\ Phys.\ {\bf A743}, 127 (2004).

     \bibitem{FKR}  K. Fukushima, C. Kouvaris, and K. Rajagopal, 
hep-ph/0408322.

\end{thebibliography}
\end{document}